\documentclass[twocolumn,showpacs,twoside,10pt,prl]{revtex4}
\usepackage{graphicx}
\usepackage{epsfig}
\usepackage{epsf}
\usepackage{amssymb}
\usepackage{amsmath}

\begin{document}

\title{A Quantum Adiabatic Algorithm for Factorization and Its Experimental Implementation}

\author{Xinhua Peng$^{1}$}
\author{Zeyang Liao$^{1}$}
\author{Nanyang Xu$^{1}$}
\author{Gan Qin$^{1}$}
\author{Xianyi Zhou$^{1}$}
\author{Dieter Suter$^{2}$}
\author{Jiangfeng Du$^{1}$}

\affiliation{Hefei National Laboratory for Physical Sciences at Microscale and
Department of Modern Physics, University of Science and Technology of China,
Hefei, Anhui 230026, People's Republic of China}
\affiliation{
Fakult\"{a}t Physik, Technische Universit\"{a}t Dortmund, 44221 Dortmund, Germany}
\date{\today}

\begin{abstract}
We propose an adiabatic quantum algorithm capable of factorizing numbers, using fewer qubits than Shor's algorithm.
We implement the algorithm in an NMR quantum information processor and experimentally factorize the number 21.
Numerical simulations indicate that the running time grows only quadratically with the number of qubits.
\end{abstract}


\maketitle

Using quantum mechanical systems as computational devices may be a
possible way to build computers that are qualitatively more powerful
than classical computers  \cite{QIQP}. The algorithms that are
adapted to the special capabilities of these devices are called
quantum algorithms. One of  the best known quantum algorithms is
Shor's algorithm for integer factorization \cite{Shor1994}. Since no
efficient factorization algorithm is known for classical computers
\cite{Knuth}, various cryptographic techniques rely on the
difficulty of finding the prime factors of large numbers
\cite{Koblitz}. However,  in 1994, Peter Shor developed a quantum
algorithm that can factorize large numbers in polynomial time
\cite{Shor1994}. This discovery was one of the main reasons for the
subsequent strong interest in quantum computation. An experimental
implementation of Shor's algorithm was demonstrated by Vandersypen
et al. \cite{Chuang}, using nuclear spins as qubits to find the
prime factors of 15. More recent experiments by Lu \textit{et al.}
\cite{Luchaoyang} and Lanyon \textit{et al.} \cite{Lanyon}  used
photons as qubits and found the same factors.

While Shor's algorithm and its experimental implementation are based
on the circuit (or network) model of quantum computing, different
models have been proposed later. Here, we consider the adiabatic
quantum computing model proposed by Farhi \emph{et al} \cite{Farhi},
The basis of this model is the quantum adiabatic theorem: A quantum
system remains in its instantaneous eigenstate if the system
Hamiltonian varies slowly enough and if there is a gap between this
eigenvalue and the rest of the Hamiltonian's spectrum \cite{Messiah,
Kato}. It has been proved to be equivalent to the conventional
circuit model \cite{Mizel}. Several adiabatic quantum algorithms have
been discussed, such as 3SAT and search of unstructured databases
\cite{Farhi, Grover, Roland}.
Compared to the network model, the adiabatic scheme appears to offer lower sentivitiy
to some perturbations and thus improved robustness against errors due to dephasing,
environmental noise and some unitary control errors \cite{Childs,Rolandrobust}.

In this paper, we propose a factorization algorithm that uses the adiabatic approach to quantum information processing.
We also implement this algorithm experimentally, using nuclear spin qubits to factorize the number 21.

There is a large class of numerical problems that can be brought into the form
of an optimization problem.
Many of them form hard problems.
The quantum adiabatic computation supplies a possible method for solving these problems.
It requires an initial Hamiltonian $H_{0}$ whose ground state $\psi_g(0)$ is well known,
and a problem Hamiltonian $H_{P}$, whose ground state encodes the solution of the optimization problem.
Implementing this method requires one to first prepare the system into the ground state of $H_0$  at $t=0$.
Subsequently, the Hamiltonian is changed, slowly enough for fulfilling the adiabatic condition,
until it is turned into the problem Hamiltonian $H_{P}$ after a time $T$.
In the simplest case, the change of the Hamiltonian is realized by an interpolation scheme
\begin{equation}
H(t)=[1-s(t)]H_{0}+s(t)H_{P}\,,
\label{e.HSys}
\end{equation}
where the function $s(t):0 \rightarrow 1$ parametrizes the interpolation.
The solution of the optimization problem is then determined by measuring the final ground state $\psi_g(T)$ of $H_{P}$.

We now apply this approach to find nontrivial prime factors of an
$\ell$-digit integer $N=p \times q$ where $p$ and $q$ are prime numbers.
Without loss of generality, we assume that $N$ is odd (in case of even $N$,
we could repeatedly divide $N$ by 2 until an odd integer is obtained).
We can write the factorization problem as an optimization problem by using the function
$f(x,y)=(N-xy)^2$, in which the variables $x$ and $y$ are positive integers \cite{Schaller}.
Clearly, the minimum of this function is reached when $x$ and $y$ are the factors of $N$.

To solve this optimization problem by the adiabatic quantum algorithm,
we must construct a problem Hamiltonian for the function $f(x,y)$,
whose ground state is the solution.
Generally, the eigenvalues of the problem Hamiltonian
are $f(x,y)$, and the corresponding eigenvectors $|x\rangle$ and $|y\rangle$ represent the variables
$x$ and $y$.
These conditions are satisfied by
\begin{eqnarray}
H_{P} = \sum_{x,y}f(x,y)|x,y \rangle\langle x,y|.
\end{eqnarray}

To determine the Hilbert space that we need for implementing this
scheme, we first consider the possible range of the variables $x$
and $y$. Since $N$ is odd, its factors $x$ and $y$ must also be odd,
i.e. its last bit is always 1 and can therefore be omitted during
the adiabatic evolution. Without loss of generality, we choose $x<y$
and $3 \leq x \leq \sqrt{N}$, $\sqrt{N} \leq y \leq \frac{N}{3}$. It
is easy to prove that $n_{x}=m(\lfloor \sqrt{N} \rfloor_{o}) -1 \leq
\lfloor \frac{\ell+1}{2} \rfloor -1$ bits are sufficient to
represent $x$ and $n_{y}=m(\lfloor \frac{N}{3} \rfloor) -1 \leq
\ell-2$ bits to represent $y$, where $\lfloor a \rfloor$ ($\lfloor a
\rfloor_{o}$) denotes the largest (odd) integer not larger than $a$,
while $m(b)$ denotes the smallest number of bits required for
representing $b$. The total number of qubits required are
$n=n_{x}+n_{y} \leq \lfloor \frac{\ell+1}{2} \rfloor +\ell-3 \sim
O(3l/2)$, which is less than the number of qubits used in Shor's
algorithm ($2\ell+1+ \lceil \log(2+\frac{1}{2\varepsilon})\rceil
\sim O(2l)$, with $\varepsilon$ the failure probability and $\lceil
c \rceil$ denotes the smallest integer not less than $c$)
\cite{Shor1994}.

Using the conventional computational basis $\{ |0\rangle,  |1\rangle \}$ = $\{ \vert \uparrow \rangle,  \vert \downarrow \rangle \}$,
it is straightforward
to construct the problem Hamiltonian:
\begin{eqnarray}
H_{P}& = &
 [NI-(2^{n_{x}-1}\frac{I-\sigma_{z}^{1}}{2}+\cdots+2^1\frac{I-\sigma_{z}^{n_{x}-1}}{2}+ I)\times
 \nonumber \\
& &
(2^{n_{y}-1}\frac{I-\sigma_{z}^{n_{x}}}{2}+\cdots+2^1\frac{I-\sigma_{z}^{n}}{2}+I)]^{2} ,
\label{e.H_P}
\end{eqnarray}
where $I$ represents the unit operator and $\sigma_{z}^{i}$ is the Pauli matrix of qubit $i$.

All computational basis states $|z_{1}z_{2}\cdots z_{n}\rangle$, with
$z_{i}=0$ or $1$, are eigenstates of $H_{P}$ and the corresponding
eigenvalues are $(N-xy)^2$, in which $x$ and $y$ are
represented by the bits $z_{1}\cdots z_{n_{x}}$ and
$z_{n_{x}+1}\cdots z_{n}$, respectively.
The lowest eigenvalue of
$H_{P}$ is $0$, and the corresponding eigenstate (i.e., the ground state) $|p'\rangle|q'\rangle$
encodes the factors $p=2p'+1$ and $q=2q'+1$.

As the initial Hamiltonian, we choose
\begin{equation}
H(0)=g(\sigma_{x}^{1}+\sigma_{x}^{2}+\cdots+\sigma_{x}^{n}) \, .
\end{equation}
This Hamiltonian describes a system, in which all the spins interact with the same magnetic field
with strength $g$, oriented along the $x$-axis.
Its ground state is
\begin{eqnarray}
|\psi_{g}(0)\rangle & = &\frac{|0\rangle-|1\rangle}{\sqrt{2}}
\otimes \frac{|0\rangle-|1\rangle}{\sqrt{2}} \otimes \cdots \otimes
\frac{|0\rangle-|1\rangle}{\sqrt{2}} \nonumber \\  & =&
\frac{1}{\sqrt{2^{n}}}\sum_{j=0}^{2^{n}}(-1)^{b(j)}|j\rangle ,
\label{Int_state}
\end{eqnarray}
where $b(j)$ is the parity of $j$ (i.e. the number of 1s in the
binary representation modulo 2). The initial state is thus an equal
superposition of all computational basis states, each representing a
combination of trial factors $x$ and $y$.

In the adiabatic process, the system evolves under the time-dependent Hamiltonian (\ref{e.HSys})
according to the Schr\"{o}dinger equation:
\begin{eqnarray}
i\frac{d}{dt}|\psi(t)\rangle=H(t)|\psi(t)\rangle ,
\label{e.Schrod}
\end{eqnarray}
with the initial condition $|\psi(0)\rangle=|\psi_{g}(0)\rangle$.
The adiabatic theorem \cite{Messiah} ensures that, if the evolution time $T$ is long enough,
the quantum system will always be close to the ground
state of $H(t)$, and the final state will be the solution of the problem.

\begin{figure}[htb]
\begin{center}
\includegraphics[width= 0.99\columnwidth]{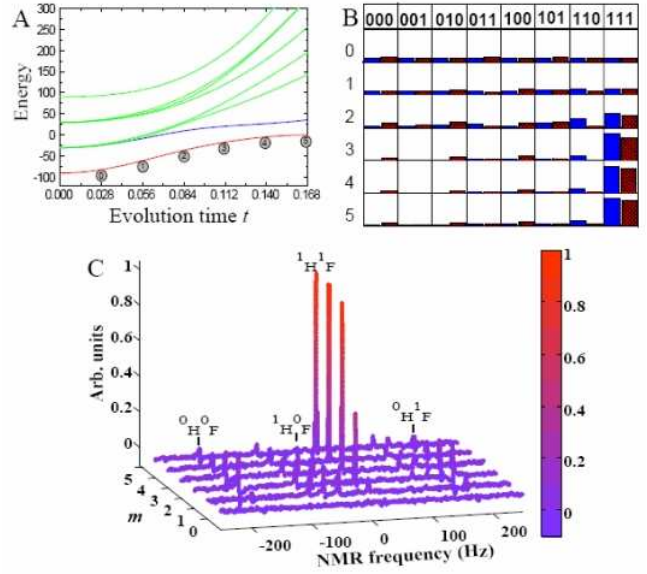}
\end{center}
\caption{(a) Energy level diagram for the adiabatic factorization of $N=21$ when $s(t)=(t/T)^2$ and $T = 0.168$.
(b) Occupation probabilities for the computational basis states
$|z_{1}z_{2}z_{3}\rangle$ for the theoretical simulation (denoted by blue bars) and experimentally reconstructed populations of the computational basis states after $m$ evolution steps (denoted by red bars). (c) Measured spectra of $^{13}$C for each adiabatic step. The four resonance lines of  $^{13}$C are labeled by the corresponding states of the two other qubits. The spectra were adjusted as absorption spectra by 180$^{o}$ phase correction, which leads to positive amplitude indicating the $\vert 1 \rangle$ subspace of the $^{13}$C qubit. A color scale indicates peak intensities, which are in arbitrary units. The system starts in an equal weight superposition and evolves to the desired final state $|111\rangle$,
which encodes the solution $p=3$, $q=7$. }
\label{Sim}
\end{figure}

As an example, we apply this algorithm to the factorization of $21$.
The number of qubits required to represent the two registers
$x$ and $y$ is $n_{x} = m(\lfloor \sqrt{21}
\rfloor_{o}) - 1=1$, $n_{y} = m(\lfloor \frac{21}{3} \rfloor)-1 = 2$.
Hence, the total number of qubits needed is $n = 3$. According to
Eq.(\ref{e.H_P}), the problem Hamiltonian is:
\begin{eqnarray}
H_{P} & = &
210I+84\sigma_{z}^{1}+88\sigma_{z}^{2}+44\sigma_{z}^{3}-20\sigma_{z}^{1}\sigma_{z}^{2}-10\sigma_{z}^{1}\sigma_{z}^{3}\nonumber\\
& &
+20\sigma_{z}^{2}\sigma_{z}^{3}-16\sigma_{z}^{1}\sigma_{z}^{2}\sigma_{z}^{3} .
\end{eqnarray}
Its energy-level diagram is shown in Fig. \ref{Sim} (a) where we
used $s(t)=(t/T)^2$ to interpolate the Hamiltonian (1), a total
evolution time $T=0.168$ and $g=30$.
Under these conditions, the
adiabatic condition is satisfied as the system evolves towards the
desired final state at $t=T$.
The blue bars in Fig. \ref{Sim} (b)
shows the numerical simulation for the evolution of the occupation
probabilities of the computational basis states
$|z_{1}z_{2}z_{3}\rangle$ when the whole adiabatic evolution is
divided into 6 equidistant steps. The ground state of the problem
Hamiltonian, $|111\rangle$ encodes the value of $x$ in the first
bit, so $p = 2 z_1 + 1 = 3$, and the value of $y$ in the second and
third bits, $q = 4 z_2 + 2 z_3 + 1 = 7$.

\begin{figure}[htb]
\begin{center}
\includegraphics[width= 0.99\columnwidth]{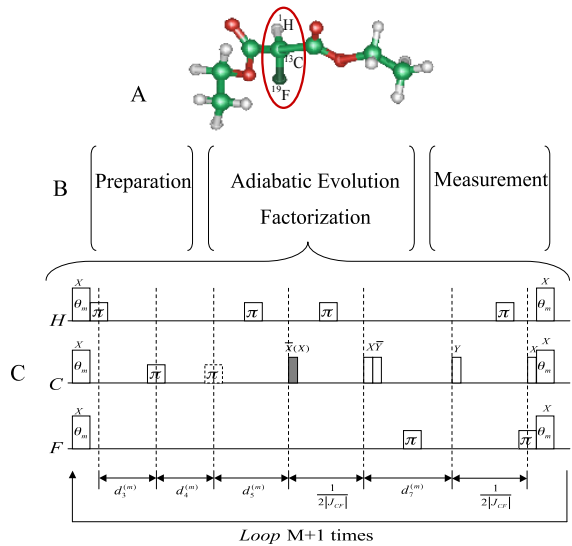}
\end{center}
\caption{ (a) Molecular structure of Diethyl-fluoromalonate, (b) schematic representation of the experiment and
(c) the pulse sequence that implements the adiabatic evolution for factorizing 21. The oval in (a) marks the
three spins used as qubits. The rectangles in (c) labelled with $\theta_m$   represent rotations by an angle $ g \left[1-\left(\frac{m}{M}\right)^r \right] \tau$,
while the narrow empty rectangles denote 90$^o$ rotations and the wide ones (labeled by $\pi$) denote the refocusing 180$^o$ pulses.
The delays are $d_3^{(m)}=10(\frac{m}{M})^r \tau(\frac{1}{\pi J_{HF}}+\frac{2}{\pi J_{HC}})$, $d_4^{(m)}=10(\frac{m}{M})^r \tau(\frac{1}{\pi J_{HF}}-\frac{2}{\pi J_{HC}})$, $d_5^{(m)}= \frac{1}{2\vert J_{CF} \vert}-20(\frac{m}{M})^r \tau(\frac{1}{\pi J_{HC}}+\frac{1}{\pi \vert J_{CF} \vert})$, and $d_7^{(m)}=32(\frac{m}{M})^r \tau\frac{1}{\pi J_{HC}}$.
Negative durations $d_5^{(m)}<0$ according to this formula were implemented as positive ones,
by omitting the dashed $\pi$ pulse preceding $d_5^{(m)}$ and replacing the the gray $-x$ pulse
immediately after the period by an $x$ pulse.}
\label{pulse}
\end{figure}

Now we turn to the real physical system (a three qubit NMR
quantum processor) to demonstrate this algorithm.
The three qubits are represented by the $^{1}$H, $^{13}$C,
and $^{19}$F nuclear spins of Diethyl-fluoromalonate.
The molecular structure is shown in Fig. \ref{pulse} (a), where the three nuclei used as qubits are marked by the oval.
The natural Hamiltonian of the three-qubit system in the rotating frame is
\begin{eqnarray}\label{HamCHF}
\mathcal{H}_{\mathit{NMR}}=  \sum_{i = 1}^{3} \frac{\omega_i}{2} \sigma^i_z + \sum_{i<j,=1}^{3} \frac{\pi J_{ij}}{2} \sigma^i_z\sigma^j_z ,
\end{eqnarray}
where $\omega_{i}$ represent the field strengths and $J_{ij}$ the coupling constants $J_{HC}=161.3$Hz, $J_{CF}=-192.2$Hz and $J_{HC}=47.6$Hz.
Experiments were carried out at room temperature, using a Bruker Avance II 500 MHz (11.7 Tesla)
spectrometer equipped with a QXI probe with pulsed field gradient.

The experiment was divided into three steps
(Fig. \ref{pulse} (b)): initial state preparation into the ground state of $H(0)$,
adiabatic passage for the time-dependent $H(t)$, and measurement of the final ground state of $H(T)$.
Starting from thermal equilibrium, we first created a pseudopure state (PPS)
\cite{Gershenfeld:1997aa,Cory:1997aa}
$\rho_{000} =\frac{1 - \epsilon}{8} \mathbf{1}+ \epsilon|000\rangle \langle 000|$,
with $\mathbf{1}$ representing the $8 \times 8$ unity operator and $\epsilon \approx 10^{-5}$  the polarization.
The ground state of $H(0)$ [Eq. (\ref{Int_state})] was prepared from $\rho_{000}$ by applying $\pi /2$ pulses along the $-y$ axis to each qubit.

The adiabatic evolution of $H(t)$ was approximated by $M+1$ discrete steps \cite{Steffen,Mitra,Peng}.
Instead of the usual linear interpolation $s(t) = t/T$, we used a polynomial interpolation, $s_m = (m/M)^r$,
with $r$ integer and $0\le m \le M$.
The unitary evolution for the discrete adiabatic passage is then
\begin{equation}
U=\prod_{m=0}^{M}U_{m}=\prod_{m=0}^{M} e^{-iH_m\tau },
\end{equation}
where the duration of each step is $\tau=T/(M+1)$.
The adiabatic limit is achieved when both $T, M \to \infty$ and $\tau \to 0$.
Using Trotter's formula, we can approximately generate the unitary operators
\begin{equation}
U_{m} \approx  e^{-i H_0(1-s_m)\frac{\tau}{2}}
e^{-i H_{P} s_m \tau} e^{-i H_0(1-s_m)\frac{\tau}{2}} + O(\tau ^3) . \nonumber
\end{equation}
The pulse sequence for the implementation of the adiabatic evolution is shown in Fig. \ref{pulse}(c).

As a suitable set of parameters, we chose the values $g=30$, $r=2$, $M=5$ and $\tau=0.028$.
This parameter set yields an adiabatic evolution that finds the solution in a relatively efficient way.
The theoretical fidelity is around 0.91.
This means that the final state has more than $90 \%$ overlap with the true solution state corresponding to the factors.

To read out the final state, only the occupation numbers of the
different computational basis states are required. To measure the
populations, we first applied a pulsed field gradient to dephase
transverse magnetization, and then a $[\pi / 2]^{i}_{-y}$ read-out
pulse to qubit $i$ and measured the resulting free induction decay
signal (FID). The readout procedure was applied to each of the three
qubits in subsequent experiments. In the experiment, we used a
sample in natural abundance, i.e. only $\approx 1 \%$ of the
molecules had a $^{13}$C nuclear spin. To distinguish those
molecules against the large background, we read out all three qubits
via the $^{13}$C channel, by applying SWAP gates and measuring the
$^{13}$C qubit.

Figure \ref{Sim} (c) shows experimental spectra obtained by reading
out the $^{13}C$ qubit, at different instances during the adiabatic
transfer. The spectrum consists of four resonance lines; in the
figure, they are labeled by the corresponding logical states of the
$^1$H and $^{19}$F qubits.
While these four resonance lines have
initially comparable amplitude, the last 2 measurements find almost
all the amplitude in the line at 15.5 Hz, which corresponds to the
$|11\rangle$ state of the $^1$H and $^{19}$F qubits. Since the
amplitude is positive, the $^{13}$C qubit must also be in the
$|1\rangle$ state.
The red bars in Fig.  \ref{Sim} b) show the
populations of the eight computational ground states. They were
obtained from a least-squares fit to the spectra measured after
reading out the three different qubits. The results confirm that the
final state has a high occupation probability on the $|111\rangle$ state, which encodes the two
factors $p=3$ and $q=7$.

These experimental results confirm that this adiabatic algorithm is
capable of factorizing numbers.
To assess its usefulness, we need to determine its time complexity.
While we could not find an analytical expression for the running time as a function
of the problem size, we performed
numerical simulations to assess its efficiency \cite{Farhi1}.
For each possible problem size, we randomly chose
$50$ different integers with nontrivial prime factors.
Then we numerically integrated the Schr\"{o}dinger equation (\ref{e.Schrod})
by a fourth-order Runge-Kutta technique.
For each run, we determined the evolution time required to reach a success probability
between 0.12 and 0.13 \cite{Farhi}.

\begin{figure}[htb]
\begin{center}
\includegraphics[width= 0.8\columnwidth]{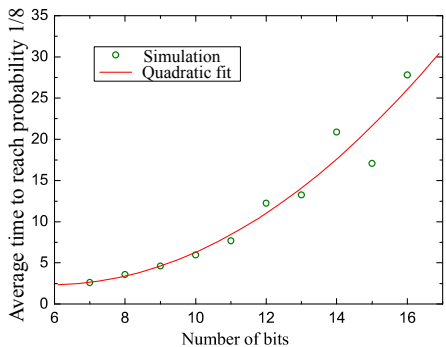}
\end{center}
\caption{Average evolution time for achieving the probability $1/8$
for $50$ instances as a function of the number of input bits $n$.
The spheres represent the simulation data for the bit number $7 \leq n \leq 16$,
while the solid line shows the quadratic fit to the data.}
\label{NuSim}
\end{figure}

In Fig. \ref{NuSim}, we plot the average of these evolution times
against the problem size.
The spheres represent simulated running times,
while the solid line is a quadratic fit to the data. The good
agreement between data points and fit suggests that the running time
grows quadratically with the size of the problem, indicating that
our algorithm scales polynomially and therefore may be efficient.

In conclusion, based on the adiabatic theorem, we propose a new
quantum algorithm for factorizing integers. Compared to Shor's
algorithm, the present method requires a significantly smaller
number of qubits.
The experimental implementation of this algorithm, using an NMR quantum
simulator, shows excellent agreement with the theoretical
expectations. To the best of our knowledge, this is the first
experimental demonstration of a quantum algorithm that factorizes an
integer larger than 15.
Furthermore, a numerical simulation of factoring up to 16-bit integers indicates that this algorithm may be efficient.

\section{Acknowledgement}

This work was supportted by National Nature Science Foundation of China, the CAS, Ministry of Education of PRC, the National Fundamental Research Program, and the DFG through Su 192/19-1.

\end{document}